\documentclass[a4paper]{article}
\usepackage{amsmath}
\def\beq{\begin{equation}}
\def\eeq{\end{equation}}
\usepackage{graphicx}
\begin{document}
\title{The Geometry of Non-Ideal Fluids}

\author{S. G. Rajeev\\s.g.rajeev@rochester.edu\\ Department of Physics and Astronomy\\ Department of Mathematics\\ University of Rochester, Rochester, NY 14618}


\maketitle

\begin{abstract}

Arnold showed that the Euler equations of an ideal fluid describe geodesics on the Lie algebra of incompressible vector fields. We generalize this to fluids with dissipation and Gaussian random forcing. The dynamics is determined by  the structure constants of a Lie algebra, along with inner products defining kinetic energy, Ohmic dissipation and the covariance of the forces. This allows us to construct tractable toy models for fluid mechanics with a finite number of degrees of freedom. We solve one of them to show how symmetries can be broken spontaneously.In another direction, we derive a deterministic equation that describes the most likely path connecting two points in the phase space of a randomly forced system: this is a WKB approximation to the Fokker-Plank-Kramer equation, analogous to the instantons of quantum theory. Applied to hydrodynamics, we derive   a  PDE system for   Navier-Stokes instantons.

\end{abstract}

\section{The Euler Equations}

The velocity of an ideal  fluid satisfies
\begin{equation}
\frac{\partial v_{i}}{\partial t}=-v^{k}\partial_{k}v_{i}-\nabla_{i}p \label{Euler}
\end{equation}
along with the auxiliary condition of incompressibility
\beq
\partial_{k}v_{k}=0.
\eeq
Arnold \cite{Arnold} advocated an elegant geometric interpretation for these equations: they describe geodesics on the Lie algebra of incompressible vector fields.

Recall that  the commutator
\beq
[u,v]^{k}=u^{j}\partial_{j}v^{k}-v^{j}\partial_{j}u^{k}.\label{incompressibleLie}
\eeq
of two incompressible (i.e., zero divergence) vector fields is still incompressible: such vector fields  form a Lie algebra. Now, any inner product on a Lie algebra defines a left-invariant  Riemannian metric on the corresponding Lie group \cite{Milnor}.
If the Lie algebra is non-abelian, this is a curved metric. But, the evolution of the tangent vectors (momenta) along a geodesic decouple from the co-ordinate variables on the group, because of the left invariance under the group action.
\subsection{Hamiltonian Formalism}
A more physical point of view is to think of the Lie algebras as defining Poisson brackets on the (angular) velocities
\beq
\left\{ v_{a},v_{b}\right\} =f_{ab}^{c}v_{c}
\eeq
The inner product defines the kinetic energy
\beq
E=\frac{1}{2}G^{ab}v_{a}v_{b}
\eeq
The resulting Hamilton's equations are  the geodesic equations of the underlying left-invariant metric on the Lie group:
\beq
\frac{dv_{a}}{dt}=\left\{ v_{a},E\right\} =f_{ab}^{c}G^{bd}v_{c}v_{d}
\eeq
More precisely, this is how the tangent vector evolves along a geodesic.

For an ideal  fluid, the kinetic energy provides a natural inner product:
\beq
E=\frac{1}{2}\int v^{2}dx \label{KE}
\eeq
Arnold's observation is that the Euler equations (\ref{Euler})  are precisely the geodesic equations defined by this metric (\ref{KE}) on the Lie algebra (\ref{incompressibleLie}) above. A more detailed derivation can be found in \cite{ChennaiFluidLectures,RajeevAnnals}.

More generally, the kinetic energy is a quadratic function of the velocities, so defines a Riemannian metric on the configuration space. In the absence of potential energy, the system moves along the geodesics of this metric: the time variable is proportional to  the arc-length.  Arnold's  point is that an ideal fluid is a conservative (hamiltonian) system without any potential energy; so it moves along the geodesics of the metric defined by its kinetic energy.  A compressible fluid has potential energy, stored in the deformation of density from its equilibrium value; so its time evolution  is not along geodesics. Also,  a viscous fluid is not conservative, so the Navier-Stokes equations are not geodesic equations either.

The curvature of a Riemannian manifold describes the stability of geodesic flows \cite{Chavel,Milnor}. If the sectional curvature $R(u,v,u,v)<0$ two geodesics that start at the same point along tangent vectors $u,v$ will diverge from each other. Arnold showed that in all except a finite number of directions, the sectional curvature is negative for the Euler flow, perhaps explaining the instability of fluid flow.

\subsection{ Example: Euler Equations of A Rigid Body }

The most well known system to which we can apply the hamiltonian formalism above is the rigid body. The Lie algebra is that of rotations
\beq
\left\{ L_{i},L_{j}\right\} =\epsilon_{ijk}L_{k}
\eeq
where $L_i$  are the components of angular momentum as measured in the comoving frame of the body.  Because this is not an inertial frame, the angular momentum may not be conserved even in the absence of external torque.
The kinetic energy is (in the frame where the moment of inertia tensor is diagonal)
\beq
\frac{1}{2}\left[G_{1}L_{1}^{2}+G_{2}L_{2}^{2}+G_{3}L_{3}^{2}\right]
\eeq
leading to the Euler equations of a rigid body. These are equations for geodesics on a triaxial ellipsoid: the group $SO(3)$ with a left-invariant metric.

 The curvatures in the principal planes are
 \beq
 K_{23}=\frac{(G_{2}-G_{3})^{2}+2G_{1}(G_{2}+G_{3})-3G_{1}^{2}}{4G_{1}G_{2}G_{3}}
 \eeq
and cyclic permutations thereof.
\subsubsection{An instability of coin flip}
For a cylinder with
\beq
h\leq\sqrt{\frac{3}{2}}r
\eeq
(e.g., coin) rotations with axis in the plane of symmetry are unstable: curvature is negative. To see this instability we would have to mark  a point of the circumference of a coin. The rotation of a coin around an axis that lies on the plane of the coin is unstable (i.e., the marked point will deviate) with respect to a perturbation of the axis to another one which also lies in the plane of the coin. An experiment to see this phenomenon is not too hard to set up with modern equipment (video cameras) but appears not to have been performed ever.

\subsection{ An Anomaly}

Recall that in hamiltonian mechanics there is always a volume form in phase space that is invariant under time evolution: this is Liouville's theorem. In our case we can ignore the position variables as they decouple from the evolution of velocities. If we simply take the divergence,
 \beq
 \frac{\partial V_{a}}{\partial v_{a}}=f_{ab}^{a}\frac{\partial E}{\partial v_{b}}.
 \eeq
If the Lie algebra is unimodular,
 \beq
 f_{ab}^{a}=0,
\eeq
this is zero
 \beq
\frac{\partial V_{a}}{\partial v_{a}}=0.
\eeq
Thus the constant  measure   is invariant.  Any semi-simple Lie algebra is unimodular. But some nilpotent and solvable ones are not.

Finite dimensional non-unimodular algebras have some other  measure $\mu$ on momentum space that is preserved by the geodesic flow:
\beq
\frac{\partial\left[\mu V_{a}\right]}{\partial v_{a}}=0.
\eeq
We will find it explicitly in some special cases below. This subtle point, that the constant measure in velocity space may not be invariant under time evolution, is reminiscent of an `anomaly' in quantum field theory. The anomalies of quantum field theory are determined by the cohomology of the Lie algebra twisted by its action on some space of fields; the  invariant  measure $\mu$ also determines an element of the (untwisted9 Lie algebra cohomology; when this element is not non-trivial we have an anomaly as well: $\mu$ is not constant. We will see that, as in field theory, such an anomaly can lead to spontaneous symmetry breaking.

\subsection{Example: The Poincare Half Plane}

The only non-abelian Lie algebra in two dimensions is the "affine Lie Algebra"
\beq
\left\{ v_{0},v_{1}\right\} =v_{1}.
\eeq
This algebra is {\em not}  unimodular. The invariant measure on velocity space is
\beq
 \mu dv=\frac{dv_{0}dv_{1}}{v_{1}}.
 \eeq
The point is that $v_0$ generates scale transformations in $v_1$, so to be invariant we must divide the naive measure $dv_0dv_1$ by $v_1$. This example therefore is a toy model where we can investigate the consequences of the anomaly we mentioned earlier.

 Any quadratic form on this Lie algebra  can be reduced to
 \beq
 E=\frac{1}{2}\left[v_{0}^{2}+v_{1}^{2}\right].
 \eeq
 The corresponding group manifold is the upper half-plane, with the Poincare\' metric. The geodesic equations are, using the above hamiltonian formalism

 \beq
 \frac{dv_{0}}{dt}=-v_{1}^{2},\ \frac{dv_{1}}{dt}=v_{0}v_{1}
 \eeq
Solutions are semi-circles in momentum space:
\beq
v_{0}=-\rho\tanh\rho t,\ v_{1}=\rho\mathrm{sech}\ \rho t.
\eeq
We will see later how dissipation and random forcing affect this dynamics.

\section{Geometry of Dissipative  Motion}

A conservative dynamical system is described by a Poisson bracket and hamiltonian (energy). Dissipation is described by (a possibly degenerate, but always non-negative)  tensor $\Gamma^{ab}$, which determines the gradient of energy:
\beq
{d\xi_a\over dt}=\{E,\xi_a\}-\Gamma_{ab}{\partial E\over \partial \xi_b}
\eeq
It is straightforward to check that the  Energy decreases:
\beq
\frac{dE}{dt}=-\Gamma_{ab}\frac{\partial E}{\partial \xi_a}\frac{\partial E}{\partial \xi_b}\leq 0
\eeq
In the cases we study in this paper, the Poisson bracket is linear ( a Lie algebra),  the energy is a quadratic function and $\Gamma_{ab}$ is constant.
\beq
\frac{dv_{a}}{dt}=-\Gamma_{ab}G^{bc}v_{c}+f_{ab}^{c}G^{bd}v_{c}v_{d}
\eeq

Even this special case is quite complex: it includes  Navier-Stokes, with the choices
 \beq
 E=\frac{1}{2}(v,Gv)=\frac{1}{2}\int v^{2}dx,\ (v,\Gamma v)=\gamma\int\left(\partial_{i}v_{j}\right)^{2}dx.
 \eeq

Thus the quadratic form of kinetic  energy is the $L^2$ norm while that of  dissipation is the $H^1$ Sobolev norm.

\section{ The Langevin And Fokker-Planck-Kramers Equations}

A standard model of randomness \cite{Kraichnan} is a  Gaussian white noise: a force whose correlations vanish fast  compared to the dynamical and dissipative time scales. If the random force is a sum of a large number of more or less independent forces,it will be a Gaussian.  The covariance of the Gaussian introduces a third positive tensor into the game.
\beq
\frac{dv_{a}}{dt}=-\Gamma_{ab}G^{bc}v_{c}+f_{ab}^{c}G^{bd}v_{c}v_{d}+\eta_{a},\ <\eta_{a}(t)\eta_{b}(t')>=D_{ab}\delta(t-t').
\eeq
This stochastic differential equation (Langevin equation) implies the Fokker-Planck-Kramers Equation for  the evolution of the probability density of velocity:

\beq
\mu\frac{\partial P}{\partial t}=\frac{\partial}{\partial v_{a}}\left[\mu\left\{ D_{ab}\frac{\partial P}{\partial v_{b}}-\left(-\Gamma_{ab}G^{bc}v_{c}+f_{ab}^{c}G^{bd}v_{c}v_{d}\right)P\right\} \right]
\eeq
 Here, $\mu$ is the invariant density of the ideal dynamics.

 The traditional derivation\cite{Kramers,ChandraStochastic}  of this FPK equation does not include the density $\mu$ : it is implicitly assumed that the invariant (Liouville) density is the standard one. In order that the total probability  be  conserved:
\beq
 {\partial\over \partial t}\int\mu Pdv=0.
 \eeq
 we must order the differentiations and insert factors of $\mu$ as above. This is one of our innovations in this paper.

 \subsection{Maxwell-Boltzmann is Equilibrium Solution for Unimodular Lie Algebras}

If the Einstein relation
\beq
\beta D_{ab}=\Gamma_{ab}
\eeq
  holds, $ P=e^{-\beta E}$ is a static solution of the FPK equations. But it may not be normalizable, because $\int\mu e^{-\beta E}dv$  might diverge. For unimodular Lie algebras, $\mu=1$ and the Maxwell-Boltzmann distribution $ P=e^{-\frac{1}{2}\beta(v,Gv)}$  is an equilibrium solution.  An example is the random motion of a rigid body. This case has been studied in detail by astronomers interested in the motion of dust grains in the interstellar medium\cite{RandomRigidBody}. We now turn to an example with $\mu\neq 1$.

 \subsection{The  Langevin Equation on the Half-Plane}

  The geodesic equations still follow for zero dissipation if we choose as hamiltonian any function $H(\rho)$ of $\rho=\sqrt{v_{0}^{2}+v_{1}^{2}.}$ We can choose $H(\rho)$  such that the FKP equation is easier to solve.  But physically, the kinetic energy would  still be $E=\frac{1}{2}\rho^{2}$. The  Fokker-Planck-Kramer reduces to
\beq
\frac{\partial Q}{\partial t}=D\frac{\partial^{2}Q}{\partial\rho^{2}}-\gamma\rho\frac{\partial Q}{\partial\rho}+\frac{D}{\rho^{2}}\frac{\partial}{\partial\theta}\left[\cosh^{2}\theta\frac{\partial Q}{\partial\theta}\right]-H'(\rho)\frac{\partial Q}{\partial\theta},\ P=e^{-\beta\frac{v^{2}}{2}}Q
\eeq
This equation is separable for $H=-\frac{k}{\rho}$. In fact the  solution can be found in closed form in terms of spheroidal functions.

Numerical solution of  toy model shows that the details of the dynamics are not important: the vestige of the dynamics that is important to the long term behavior is the invariant measure $ \mu$. We can even set $H=0$ and get a qualitatively similar equilibrium distribution. In this case the solution is quite simple\cite{AbrSt}:
\beq
P(v_{0},v_{1})\frac{dv_{0}dv_{1}}{v_{1}}=\frac{v_{1}}{\rho^{2}}e^{-\frac{\beta}{2}\rho^{2}}
\left[1-\frac{1}{\sqrt{\beta}\rho}
e^{-\frac{\beta\rho^{2}}{4}}\mathrm{erf}\left(\frac{\sqrt{\beta}\rho}{2}\right)\right]dv_{0}dv_{1}
\eeq
The main point is that this solution does not have a peak at minimum energy $v_0=v_1=0$; instead the peak is at some finite value of $v_1$. The anomaly drives a spontaneous breaking of the rotation invariance in the energy $E={1\over 2}[v_0^2+v_1^2]$. Even if the two components of velocity appear symmetrically in energy, their commutation relations are different. These commutation relation determine the measure of integration which is singular at $v_1=0$. It is this singularity that pushes the most likely equilibrium value of $v_1$ away from zero.

Thus we see that anomaly we discovered can drive a spontaneous symmetry breakdown: a steady state flow along the $x_1$ axis is generated in the equilibrium. Perhaps this is a toy model for the formation of  flows such as hurricanes and ocean currents  which persist for long times despite the instability of the underlying fluid system.

\section{The WKB Approximation to the FPK Equation}

It is often important to know the probability that a certain initial condition will evolve to a given final condition under the influence of random forces. It might be important to compute this even when this probability is small: there could be final states that are catastrophic so that even a small chance of them  happening cannot be ignored. If the probability is small, we can use the WKB approximation as for the Schrodinger equation;  put  $P=e^{-\Phi }$ and ignore second derivatives of $\Phi$:
\beq
\frac{\partial\Phi}{\partial t}+D_{ab}\frac{\partial\Phi}{\partial v_{a}}\frac{\partial\Phi}{\partial v_{b}}+\left(-\Gamma_{ab}G^{bc}v_{c}+f_{ab}^{c}G^{bd}v_{c}v_{d}\right)\frac{\partial\Phi}{\partial v_{a}}=0
\eeq
These are the Hamilton-Jacobi equations of the hamiltonian
\beq
H=D_{ab}w^{a}w^{b}+\left(-\Gamma_{ab}G^{bc}v_{c}+f_{ab}^{c}G^{bd}v_{c}v_{d}\right)w^{a},
\eeq
where $w_a$ are the canonical conjugates of the variables $v_a$:
\beq
\ \left\{ v_{a},v_{b}\right\} _{1}=0=\left\{ w^{a},w^{b}\right\} ,\ \left\{ v_{a}w^{b}\right\} _{1}=\delta_{a}^{b}
\eeq
Kramer used  this WKB approximation to the FPK  in chemical kinetics; we extend them to a more general framework. In quantum mechanics, the WKB approximation works when the quantum tunneling amplitudes  are small; the path with the greatest probability is called the instanton. The analogue in our case is the solution to the Hamilton's equations following from the above efectie hamiltonian.
\beq
\frac{dv_{a}}{dt}=2D_{ab}w^{b}-\Gamma_{ab}G^{bc}v_{c}+f_{ab}^{c}G^{bd}v_{c}v_{d},\ \frac{dw^{c}}{dt}=\left(\Gamma_{ab}G^{bc}-2f_{ab}^{c}G^{bd}v_{d}\right)w^{a}
\eeq

 Notice that this effective hamiltonian mechanics of rare transitions has twice the number of degrees of freedom as the original conservative dynamics we started with: we can specify initial and final conditions independently. The action of the path that solves this equation with given initial and final states gives the log of the transition probability:
 \beq
 \Phi(v',v,T)=\int_0^T[w^a\dot{v_a}-H(v,w)]dt
 \eeq
 \subsection{Instantons of Random Motion on the Half-Plane}
With an isotropic dissipation tensor, these instanton equations are quite simple for the half-plane:
\beq
\frac{dv_{0}}{dt}=w_{0}-\gamma v_{0}-v_{1}^{2},\ \frac{dv_{1}}{dt}=w_{1}-\gamma v_{1}+v_{0}v_{1}
\eeq
\beq
\frac{dw_{0}}{dt}=\gamma w_{0}-v_{1}w_{1},\ \frac{dw_{1}}{dt}=\gamma w_{1}+2v_{1}w_{0}-v_{0}w_{1}
\eeq
We could not find the general solution to this system.  But notice the special solution with $w_{a}=2\gamma v_{a}$. This is just the time reverse of the dissipative solution!
 This solution leads to the Maxwell-Boltzmann  distribution: agrees with the exact solution for large energies. WKB is not good for small energies, where the exact answer departs from MB distribution. This is the phenomenon of spontaneous symmetry breakdown we discovered in the last section.

 \subsection{Navier-Stokes Instantons }

  We can now derive the PDE describing the most likely time evolution of a randomly forced dissipative fluid:
  \beq
\frac{\partial v_{i}}{\partial t}=F_{i}+\gamma\partial^{2}v_{i}-v_{j}\partial_{j}v_{i}-\partial_{i}p,\quad\partial_{k}v_{k}=0.
\eeq
\beq
\frac{\partial w_{i}}{\partial t}=-\gamma\partial^{2}w_{i}+w_{j}\partial_{j}v_{i}+v_{j}\partial_{j}w_{i}-\partial_{i}\phi,\ \partial_{k}w_{k}=0.
\eeq
The covariance of the fluctuations is likely to be an integral operator\cite{Kraichnan}
\beq
D_{ij}(x,x')=\int e^{2\pi ik\cdot(x-x')}\left[\delta_{ij}-\frac{k_{i}k_{j}}{k^{2}}\right]\tilde{D}(k)dk.
\eeq
If we choose
\beq
\tilde{D}(k)=|k|^{-2}
\eeq
we get an auxiliary  differential equation to complete the above system
\beq
\partial^{2}F_{i}=2w_{i}.
\eeq
With this choice, the random forcing grows with spatial scale, while dissipation is more important at short scales: what is expected in fluid mechanics. The above equation should be a good approximation to calculate probabilities and paths of large deviations from Navier-Stokes.  Such rare events are important in many situations: the path of a hurricane, or the chances of a tsunami or a rogue wave. Similar instanton equations can be derived for Rotating Shallow Water equations, the Stochastic Loewner Evolution\cite{SLE}, two dimensional incompressible flow etc. We hope to return to these questions in later publications.  Numerical solution of our system is not much harder than the solution of Navier-Stokes itself: only twice as many variables; moreover, the new variables appear linearly.

\section*{Acknowledgement}
I thank M. Gordina, G. Krishnaswamy, V.V. Sreedhar and C. Mueller  for discussions. This work was supported in part by a grant from the
US Department of Energy under contract DE-FG02-91ER40685.

\pagebreak

\end{document}